\documentclass[12pt]{article}
\usepackage[english]{babel}
\usepackage{amsthm}
\usepackage{amsfonts, mathtools}
\usepackage{graphicx}
\usepackage[usenames,dvipsnames]{color}
\usepackage{subfigure}
\usepackage[right,pagewise,displaymath, mathlines]{lineno}
\usepackage{epstopdf}
\usepackage{color}

\usepackage{rotating}
\usepackage{amssymb,amsmath}
\usepackage{fullpage,enumerate,subfig,multirow}
\usepackage{fancybox,enumerate}

\numberwithin{equation}{section}
\numberwithin{figure}{section}
\numberwithin{table}{section}

\numberwithin{equation}{section}
\begin{document}

\begin{center}
{\bf  On classical and Bayesian inference for bivariate  Poisson conditionals distributions: Theory, methods and applications}
\end{center}

\bigskip
\begin{center}
{   Barry C. Arnold$^{1},$  Indranil Ghosh$^{2},$\\
{\small $^{1}${ University of California, Riverside }\\
 $^{2}${University of North Carolina, Wilmington, USA}\\
  }
}
\end{center}

\bigskip

\begin{abstract}
Bivariate count data arise in several different disciplines (epidemiology, marketing, sports statistics, etc., to name but a few) and the bivariate Poisson distribution which is a generalization of the Poisson distribution plays an important role in modeling such data. In this article, we consider the inferential aspect of a bivariate Poisson conditionals distribution for which both the conditionals are Poisson but the marginals are typically non-Poisson. It has Poisson marginals only in the case of independence. It appears that a simple iterative procedure under the maximum likelihood method performs quite well as compared with other numerical subroutines, as one would expect in such a case where the MLEs are not available in closed form. In the Bayesian paradigm, both conjugate priors and non-conjugate priors have been utilized and a comparison study has been made via a simulation study. For illustrative purposes, a real-life data set is re-analyzed to exhibit the utility of the proposed two methods of estimation, one under the frequentist approach and the other under the Bayesian paradigm.

\end{abstract}


\section{Introduction}
Bivariate count data arise in many circumstances. For example, in medicine, we may have pretreatment and post treatment measurements of the same individuals, or we may consider the incidence of two diseases  in certain sites. Paired count data also arise  in various other domains affecting our daily lives, such as economics, medical science, sports medicine, reliability of a production process  etc.
Bivariate discrete Poisson distributions have enjoyed a good amount of attention over the last couple of decades or so. Various different versions of the bivariate Poisson distribution have been adequately discussed in the literature. Additionally, several different strategies to estimate the model parameters under both the frequentist as well as under the Bayesian paradigm have been developed.

For a comprehensive treatment of the bivariate Poisson distribution and its multivariate extensions the reader can refer to Kocherlakota and Kocherlakota (2017), and Johnson, Kotz, and Balakrishnan (1997). Below, we provide a non-exhaustive list of related pertinent references.

Recently, to remedy against the problem of   computational difficulties related to statistical inference for a bivariate and multivariate Poisson distribution, many authors 
 have proposed efficient and tricky strategies. Some useful references in this context can be cited as follows.  For example,  the even-points method by  Papageorgiou and Kemp (1977) in the context of a bivariate generalized Poisson distribution; the  use of conditional-even points method introduced by Papageorgiou and Loukas (1988) to estimate the model parameters of  a bivariate Poisson distribution can be cited as well.
Holgate (1964) discussed the estimation of the covariance parameter for a correlated bivariate Poisson distribution and advocated for the use of iterative method of solving the likelihood equations as compared to the method of moments strategy under the classical set-up. Belov (1993) has established the result on the uniqueness of the maximum likelihood estimates for the parameters of the bivariate Poisson distribution. For a Monte Carlo study concerning the performance of alternative estimators, see Paul and Ho (1989).\\
 Estimation of parameters under the Bayesian paradigm has also been developed for univariate, bivariate and multivariate Poisson distributions. For example, Karlis and Ntzoufras (2006) have discussed the Bayesian analysis of the difference of count data assuming a bivariate Poisson distribution. Karlis and Tsiamyrtzis (2008) provided a framework to conduct an exact Bayesian analysis for bivariate Poisson data assuming conjugate gamma priors. Mo and Kockelman (2006) developed a Bayesian multivariate Poisson regression model useful in modeling injury count data. Tsionas (1999) has discussed the Bayesian analysis of the multivariate Poisson distribution based on Gibbs sampling and by invoking a data augmentation strategy.\\

 However, there has been not much discussion and study of   negatively correlated bivariate Poisson distributions. Consequently, not much work has been done regarding Bayesian estimating of the model parameters. This serves as one of the major motivations to carry out the present research work. 

In this article,  we discuss the estimation (under both the frequentist and the Bayesian paradigm) of the model parameters of a bivariate Poisson conditionals distribution independently discussed by Obrechkoff (1963) and in Arnold et al. (1999). This distribution has also been independently discussed by Wesolowski (1996).
The rest of this paper is organized as follows. In Section $2,$  we introduce the bivariate Poisson conditionals distributions due to Arnold et al. (1999) and  Obrechkoff.
In Section $3,$ we discuss the maximum likelihood method of estimating the model parameters via an iterative process that is different from that used by Ghosh et al. (2021). 
Section $5$ outlines Bayesian inference for the bivariate Poisson conditional type distributions  using informative priors. 
The simulated results are presented in Section $6.$ Section $7$ discusses the Bayesian estimation using the posterior mode(s) as the posterior summary for the model parameters. For illustrative purposes, a real-life data set has been re-analyzed to exhibit the efficacy of the proposed two methods of estimation under the frequentist and under the Bayesian framework in Section $8.$
Finally, some concluding remarks are provided in Section $9.$

\section{Bivariate Poisson conditionals distributions}
 We begin our discussion in this section by introducing a bivariate discrete distribution for which both sets of  conditionals are  univariate Poisson according to Arnold et al. (1999) (p.96-97). This probability model also appears in Obrechkoff (1963) and in Wesolowski (1996).
 
Let us assume the following:

\begin{itemize}
\item $X|Y=y\sim Poisson\left(\lambda_{1}\lambda^{y}_{3}\right),$  for each fixed $Y=y.$

\item $Y|X=x\sim Poisson\left(\lambda_{2}\lambda^{x}_{3}\right),$ for each fixed $X=x.$
\end{itemize}

\noindent Here, $\left(\lambda_{1}, \lambda_{2}\right)> 0, \quad 0 <\lambda_{3}\leq 1.$
Note that if $\lambda_{3}=1,$  $X$ and $Y$ are independent.

According to Arnold et al. (1999) [see Theorem $4.1,$ page $76$] the associated joint p.m.f. will be

\begin{equation} \label{BPC-dens}
P\left(X=x, Y=y\right)
=K\left(\lambda_{1},\lambda_{2},\lambda_{3}\right) \displaystyle\times \frac{\lambda^{x}_{1}\lambda^{y}_{2}\lambda^{xy}_{3}}{x!y!},
\end{equation}

\noindent where $x=0,1,2,\cdots; \quad y=0,1,2,\cdots,$ and $K\left(\lambda_{1},\lambda_{2},\lambda_{3}\right)$ is the normalizing constant and

\begin{equation*}
K^{-1}=K^{-1}\left(\lambda_{1}, \lambda_{2},\lambda_{3}\right)
=\sum_{y=0}^{\infty}\frac{\lambda^{y}_{2}}{y!}\exp\left(\lambda_{1}\lambda^{y}_{3}\right)
=\sum_{x=0}^{\infty} \frac{\lambda^{x}_{1}}{x!}\exp\left(\lambda_{2}\lambda^{x}_{3}\right).
\end{equation*}

The  general assumption of Poisson conditionals forces one to have this structure.

We will denote the bivariate  Poisson distribution  of  the pair $\left(X,Y\right)$  with the p.m.f. in (\ref{BPC-dens})  as $BPC\left(\lambda_{1},\lambda_{2},\lambda_{3}\right).$   Several useful structural properties of the joint p.m.f. in (\ref{BPC-dens}) have been discussed in Ghosh et al. (2021).

In the next section we will focus  our attention on the maximum likelihood estimation of the model parameters for the BPC distribution in (\ref{BPC-dens}) via a simple iterative strategy. The adopted strategy is different from the approach used  in Ghosh et al. (2021). In that paper, the authors discussed maximum likelihood estimation using a copula based approach.

\section{Iterative maximum likelihood estimation for the BPC distribution}
For a random sample of size $n,$ the log-likelihood function of the bivariate Poisson conditionals distribution will be given by

\begin{equation}
\ell(\underline{\lambda})=-n \log J(\underline{\lambda})+t_1 \log \lambda_1 +t_2 \log \lambda_2+t_3 \log \lambda_3-\sum_{i=1}^n \log (x_i!) -\sum_{i=1}^n \log (y_i!),
\end{equation}

\noindent where

$$J(\lambda_1,\lambda_2,\lambda_3)=\sum_{y=0}^{\infty}\frac{\lambda_2^y}{y!}exp\{\lambda_1\lambda_3^y\}=\sum_{x=0}^{\infty}\frac{\lambda_1^x}{x!}exp\{\lambda_2\lambda_3^x\}.$$

From  Eq. (3.1), the MLEs are obtained by taking partial derivatives w.r.t. $\lambda_1, \lambda_2, \lambda_3$ and setting them equal to zero.

\begin{eqnarray}
\frac{\partial \ell(\underline{\lambda})}{\partial \lambda_1}=-\frac{n}{J(\underline{\lambda})}\left[\frac{\partial J(\lambda_1,\lambda_2,\lambda_3)}{\partial \lambda_1}\right]+\frac{t_1}{\lambda_1},\\
\frac{\partial \ell(\underline{\lambda})}{\partial \lambda_2}
=-\frac{n}{J(\underline{\lambda})}\left[\frac{\partial J(\lambda_1,\lambda_2,\lambda_3)}{\partial \lambda_2}\right]+\frac{t_2}{\lambda_2},\\
\frac{\partial \ell(\underline{\lambda})}{\partial \lambda_3}
=-\frac{n}{J(\underline{\lambda})}\left[\frac{\partial J(\lambda_1,\lambda_2,\lambda_3)}{\partial \lambda_3}\right]+\frac{t_3}{\lambda_3},
\end{eqnarray}

\noindent where $t_{1}=\sum_{i=1}^{n}x_{i}, \quad t_{2}=\sum_{i=1}^{n}y_{i}, \quad t_{3}=\sum_{i=1}^{n}x_{i}y_{i}.$

\noindent Because of the nature of the $J(\lambda_1,\lambda_2,\lambda_3)$  function,  we can rewrite the  Eqs. (3.2)-(3.4) as

\begin{eqnarray*}
\frac{J(\lambda_1,\lambda_2\lambda_3,\lambda_3)}{J(\lambda_1,\lambda_2,\lambda_3)}=\frac{t_1}{n \lambda_1},\\
\\
\frac{J(\lambda_1 \lambda_3,\lambda_2,\lambda_3)}{J(\lambda_1,\lambda_2,\lambda_3)}=\frac{t_2}{n \lambda_2}, \\
\\
\frac{\lambda_1 \lambda_2 J(\lambda_1 \lambda_3, \lambda_2 \lambda_3,\lambda_3)}{J(\lambda_1,\lambda_2,\lambda_3)}=\frac{t_3}{n \lambda_3}.
\end{eqnarray*}

\noindent It can be easily verified that the asymptotic variance-covariance of the MLEs of $\lambda_{1}, \lambda_{2},$ and $\lambda_{3}$ cannot be obtained  analytically because of the complicated nature of the expectations.

Therefore,  we obtain the approximate asymptotic variance-covariance matrix for the MLEs by getting the inverse of the observed FIM,  which is as follows.

\begin{eqnarray}
I\left(\widehat{\lambda_1},\widehat{\lambda_2}, \widehat{\lambda_3}\right)&=&
\begin{bmatrix}
-\frac{\partial^{2}\ell(\underline{\lambda})}{\partial\lambda^{2}_1}& -\frac{\partial^{2}\ell(\underline{\lambda})}{\partial\lambda_1\partial\lambda_2}
&-\frac{\partial^{2}\ell(\underline{\lambda})}{\partial\lambda_1\partial\lambda_2}\\
\frac{\partial^{2}\ell(\underline{\lambda})}{\partial\lambda_2\partial\lambda_1}&-\frac{\partial^{2}\ell(\underline{\lambda})}{\partial\lambda^{2}_2}&
-\frac{\partial^{2}\ell(\underline{\lambda})}{\partial\lambda_2\partial\lambda_3}\\
-\frac{\partial^{2}\ell(\underline{\lambda})}{\partial\lambda_3\partial\lambda_1}&-\frac{\partial^{2}\ell(\underline{\lambda})}{\partial\lambda_3\partial\lambda_2}& -\frac{\partial^{2}J(\underline{\lambda})}{\partial\lambda^{3}_3}
\end{bmatrix}\notag\\
&=& \begin{bmatrix}
Var(\widehat{\lambda_1}) & &\\
& Var(\widehat{\lambda_2}) & \\
& & Var(\widehat{\lambda_3})
\end{bmatrix}.
\end{eqnarray}

\noindent The asymptotic variance-covariance matrix of the MLE $\underline\lambda = ({\hat \lambda }_{1}, {\hat \lambda }_{2}, {\hat \lambda }_{3})$ can be obtained from the inverse of the observed Fisher information matrix as
$$ {\bf V} = {\bf I}^{-1}(\underline\lambda) \stackrel{def.}{=}
\begin{bmatrix}
v_{11} & v_{12} & v_{13}\\
       & v_{22} & v_{23}\\
       &        & v_{33}.
  \end{bmatrix}
$$

\noindent Under mild regularity conditions, $$\left(\widehat{\lambda_1}, \widehat{\lambda_2}, \widehat{\lambda_3}\right)\sim N_{3}\bigg((\lambda_1,\lambda_2,\lambda_3), {\bf V}\bigg).$$

Therefore, a $100\left(1-\tau\right)$\% approximate confidence intervals of the parameters $\widehat{\lambda_i}$ will be

$$\widehat{\lambda_i}\pm Z\left(1-\tau/2\right) \times \sqrt{v_{ii}}, $$  $i=1,2,3,$ where $Z_{q}$ is the 100$q$-th upper percentile of the standard normal distribution.

\noindent Next, in this case, the elements of of the observed FIM are:

\begin{itemize}
\item $\frac{\partial^{2}\ell(\underline{\lambda})}{\partial\lambda^{2}_1}=-\frac{t_1}{\lambda^{2}_{1}}-n
\bigg[\frac{J(\underline{\lambda})\times J(\lambda_1,\lambda_2\lambda^{2}_3,\lambda_3)-\left(J(\lambda_1,\lambda_2\lambda_3,\lambda_3)\right)^{2}}{J^{2}(\underline{\lambda})}\bigg].$

\item $\frac{\partial^{2}\ell(\underline{\lambda})}{\partial\lambda^{2}_2}=-\frac{t_2}{\lambda^{2}_{2}}-n
\bigg[\frac{J(\underline{\lambda})\times J(\lambda_1\lambda^{2}_3,\lambda_2,\lambda_3)-\left(J(\lambda_1\lambda_3,\lambda_2,\lambda_3)\right)^{2}}{J^{2}(\underline{\lambda})}\bigg].$

\item  $\frac{\partial^{2}\ell(\underline{\lambda})}{\partial\lambda^{2}_3}=-\frac{t_3}{\lambda^{2}_{3}}-n
\bigg[\frac{J(\underline{\lambda})\times J(\lambda_1\lambda^{2}_3,\lambda_2\lambda^{2}_3,\lambda_3)\times \left(\lambda_1\lambda_2\right)^{2}-\left(\lambda_1\lambda_2\right)\times\left(J(\lambda_1\lambda_3,\lambda_2\lambda_3,\lambda_3)\right)^{2}}{J^{2}(\underline{\lambda})}\bigg].$

\item Again, 

\begin{equation*}
\frac{\partial^{2}\ell(\underline{\lambda})}{\partial\lambda_1\partial\lambda_2}
=J\left(\lambda_1\lambda_3,\lambda_2\lambda^{2}_3,\lambda_3\right).
\end{equation*}

\item Again, 

\begin{equation*}
\frac{\partial^{2}\ell(\underline{\lambda})}{\partial\lambda_1\partial\lambda_3}
=\lambda_{2} J\left(\lambda_1\lambda_3,\lambda_2\lambda_3,\lambda_3\right)+\left(\lambda_1\lambda_3\right)J\left(\lambda_1\lambda_3,\lambda_2\lambda^{2}_3,\lambda_3\right).
\end{equation*}

\item Also,

\begin{equation*}
\frac{\partial^{2}\ell(\underline{\lambda})}{\partial\lambda_2\partial\lambda_3}
=\lambda_{3} J\left(\lambda_1\lambda^{2}_3,\lambda_2\lambda_3,\lambda_3\right).
\end{equation*}

\end{itemize}

Instead of using any optimization program/subroutine, we consider the following  approach to obtain the MLEs of  $\lambda_1,\lambda_2, \lambda_3$. Observe that, the above likelihood equations  can be re-written as

\begin{eqnarray*}
\lambda_1=\frac{t_1J(\lambda_1,\lambda_2,\lambda_3)}{nJ(\lambda_1,\lambda_2\lambda_3,\lambda_3)} \hspace{1in},(A)\\
\\
\lambda_2=\frac{t_2J(\lambda_1,\lambda_2,\lambda_3)}{nJ(\lambda_1 \lambda_3, \lambda_2, \lambda_3)} ,\hspace{1in}(B)\\
\\
\lambda_3=\frac{t_3J(\lambda_1,\lambda_2,\lambda_3)}{n\lambda_1 \lambda_2 J(\lambda_1 \lambda_3, \lambda_2 \lambda_3,\lambda_3)}. \hspace{1in}(C)
\end{eqnarray*}

\bigskip

\noindent Next, we adopt the following simple (repetitive) process:

\begin{itemize}
\item First, we pick initial values for the the three $\lambda_{i}$'s.

\item Next,  use (A) with the three current values for the $\lambda$'s on the right side to update $\lambda_1.$

\item Next, use (B)  with the three current values for the $\lambda$'s on the right side to update $\lambda_2.$

\item Finally,  use (C)  with the three current values for the $\lambda$'s on the right side  to update  $\lambda_3.$

\item We continue this process until the process converges in the sense that we stop at stage $m$ if $|\lambda^{m}_{i}-\lambda^{m+1}_{i}|<\epsilon,$ where $\epsilon$ is a very small quantity $<0.005.$

\end{itemize}

\section{Simulation Study}
 Let us assume that a random sample of size $n$ is drawn from the joint p.m.f. in  (2.1).
 In particular, we consider the sample sizes $n$ = 50, 75 and 100 with the following four sets of choices of the model parameters:

\begin{itemize}
\item[(a)] Choice 1: $\lambda_{1}= 2$, $\lambda_{2}=2.5$ and  $\lambda_{3}=0.35.$
\item[(b)] Choice 2: $\lambda_{1}= 1.75$, $\lambda_{2}=3.25$ and  $\lambda_{3}=0.45.$
\item[(c)] Choice 3: $\lambda_{1}= 2.5$, $\lambda_{2}=1.5$ and  $\lambda_{3}=0.55.$
\item[(d)] Choice 4: $\lambda_{1}= 3.5$, $\lambda_{2}=4$ and  $\lambda_{3}=0.75.$
\end{itemize}
 Random samples from the BPC distribution are generated using the techniques  discussed in Shin and Pasupathy (2010). The MLEs of $\lambda_{1}$, $\lambda_{2},$ and $\lambda_{3}$ are obtained by adopting the strategy  described in the previous section.
 
 For each of these choices above, the following initial values of the parameters are considered 

\begin{itemize}
\item[(a)] Initial values for Choice 1: $\lambda_{1}= 1.04$, $\lambda_{2}=1.23$ and  $\lambda_{3}=0.125.$
\item[(b)] Initial values for Choice 2: $\lambda_{1}= 0.98$, $\lambda_{2}=1.46$ and  $\lambda_{3}=0.27.$
\item[(c)] Initial values for Choice 3: $\lambda_{1}= 1.12$, $\lambda_{2}=1.03$ and  $\lambda_{3}=0.28.$
\item[(d)] Initial values for Choice 4: $\lambda_{1}= 1.74$, $\lambda_{2}=2.23$ and  $\lambda_{3}=0.18.$
\end{itemize}

\begin{sidewaystable}
\centering
\small
\caption{Simulated coverage probabilities (CP) and average widths (AW) of the MLEs of the parameters in the BPCN distribution for various choices of $\underline\lambda$}
\label{biasmse}
\begin{tabular}{c c c c c c c c c  c c c c c c} \hline
            Parameter choice    & \multicolumn{7}{c}{Based on asymptotic variances from inverting ${\bf I}(\underline\lambda)$} & \multicolumn{6}{c}{Based on bootstrap variances} \\
              & \multicolumn{2}{c}{$\lambda_{1}$} &  \multicolumn{2}{c}{$\lambda_{2}$}  & \multicolumn{2}{c}{$\lambda_{3}$} & \% of negative & \multicolumn{2}{c}{$\lambda_{1}$} &  \multicolumn{2}{c}{$\lambda_{2}$}  & \multicolumn{2}{c}{$\lambda_{3}$}\\
  & $n$ & CP & AW & CP & AW  & CP & AW  & variances &  CP & AW & CP & AW  & CP & AW  \\ \hline
  Choice 1& 50  & 0.950 & 1.377 & 0.952 & 0.563 & 0.992 & 2.361 & 0.0475  & 0.913 & 1.388 & 0.938 & 0.535 & 0.922 & 1.732\\
      &75  & 0.939 & 1.243 & 0.953 & 0.487 & 0.986 & 1.261 & 0.090  & 0.814 & 1.344 & 0.938 & 0.485 & 0.917 & 1.534\\
     & 100 & 0.935 & 1.137 & 0.959 & 0.432 & 0.982 & 0.584 & 0.120  & 0.923 & 1.299 & 0.935 & 0.452 & 0.927 & 1.335 \\ \hline
 Choice 2&50  & 0.905 & 1.444 & 0.940 & 0.577 & 0.997 & 2.569 & 0.130  & 0.912 & 1.508 & 0.950 & 0.570 & 0.958 & 1.137 \\
   &   75  & 0.882 & 1.292 & 0.943 & 0.498 & 0.993 & 1.189 & 0.170  & 0.940 & 1.422 & 0.945 & 0.508 & 0.959 & 1.032 \\
     & 100 & 0.853 & 1.203 & 0.943 & 0.448 & 0.988 & 1.032 & 0.090  & 0.925 & 1.362 & 0.944 & 0.469 & 0.954 & 0.812 \\ \hline
 Choice 3 &50  & 0.950 & 1.371 & 0.949 & 0.561 & 0.993 & 2.481 & 0.110  & 0.918 & 1.392 & 0.943 & 0.535 & 0.926 & 1.643 \\
      &75  & 0.941 & 1.229 & 0.956 & 0.484 & 0.986 & 1.264 & 0.170  & 0.904 & 1.345 & 0.935 & 0.486 & 0.914 & 1.345 \\
      &100 & 0.930 & 1.117 & 0.955 & 0.428 & 0.978 & 1.172 & 0.110  & 0.911 & 1.292 & 0.934 & 0.449 & 0.924 & 0.733 \\ \hline
 Choice 4&50  & 0.904 & 1.437 & 0.940 & 0.575 & 0.997 & 2.556 & 0.130  & 0.936 & 1.478 & 0.945 & 0.564 & 0.955 & 1.542 \\
      &75  & 0.882 & 1.289 & 0.943 & 0.496 & 0.993 & 1.218 & 0.090  & 0.935 & 1.406 & 0.944 & 0.504 & 0.953 & 1.046 \\
    &100 & 0.853 & 1.199 & 0.943 & 0.447 & 0.988 & 1.043 & 0.100  & 0.921 & 1.373 & 0.947 & 0.474 & 0.947 & 0.938 \\ \hline
 \end{tabular}
\end{sidewaystable}

One may observe from  Table $4.1,$   that the estimated MSEs for the three parameters $\lambda_{1}$, $\lambda_{2}$ and  $\lambda_{3}$ decrease as the sample size increases. However, for the estimated biases, there is not a steady decreasing pattern with the increase of sample sizes, and   on the contrary, in some cases, it appears that there is a negligible amount (by $0.01-0.05$) of increase.
We observe that the direction of the estimated biases  of  the MLE of $\lambda_{3}$ is the same as the sign of the true value of the parameter $\lambda_{3}$. Moreover, the estimated MSEs of $\lambda_{3}$ is larger than the MSEs of $\lambda_{1}$ and $\lambda_{2}$.

Additionally, from Table $4.1,$ one may  also observe the following

\begin{itemize}
 \item that the proportions of cases in which negative variance estimates are obtained is negligibly small.  
 
 \item Additionally, the computed  approximate confidence intervals based on bootstrap variances performs satisfactorily well. Note that these approximate confidence intervals can be used as an  alternative when the asymptotic variances are negative (for pertinent details, see Ghosh and Ng (2019) and the references cited therein). 
  \end{itemize}

\section{Bayesian inference}
 Since the classical methods of estimation for bivariate discrete probability models does not always yield satisfactory results due to several factors, such as likelihood involving ubiquitous normalizing constants, non-existence of efficient algorithms to obtain global maximums for the model parameter(s) as opposed to local maximums, etc., it is legitimate to consider a Bayesian approach in this context. There are several advantages of conducting a Bayesian analysis, especially for bivariate discrete probability models (for pertinent details, see Bermúdez, L., \& Karlis, D. (2011).  In this section, we begin our discussion on the Bayesian estimation by assuming the conjugate prior set-up at first for the joint p.m.f. as given in Eq. (2.1). In this case, we are dealing with a three parameter exponential family, so a conjugate prior will exist.  First we reparametrize by defining new parameters as follows. 

$$\delta_i=\log \lambda_i, \ \ i=1,2,3.$$
 
Note that $\delta_1,\delta_2 \in (-\infty,\infty)$ while $\delta_3 \in (-\infty,0].$ 

The BPC joint p.m.f. in Eq. (2.1) can be re-written as

\begin{equation}
f\left(x,y;\vec{\delta}\right)=\frac{\widetilde{K}(\delta_1,\delta_2,\delta_3) \exp[\delta_1 x+\delta_2 y+\delta_3 xy]}{x!y!},
\end{equation}

\noindent where $x$ and $y$ are non-negative integers, and $\vec{\delta}=\left(\delta_1,\delta_2,\delta_3\right).$

The  associated likelihood function corresponding to a sample of size $n$ will then be

\begin{equation}
L(\underline{\delta})=\frac{[ \widetilde{K}(\delta_1,\delta_2,\delta_3)]^n \exp[\delta_1 \sum x_i+\delta_2 \sum y_i+\delta_3 \sum x_iy_i]}{\prod x_i ! \prod y_i !}.
\end{equation}

As a conjugate prior, one may consider the following
\begin{equation}
f_{\underline{\eta}}(\underline{\delta}) \propto [ \widetilde{K}(\delta_1,\delta_2,\delta_3)]^{\eta_0} \exp[\eta_1 \delta_1 +\eta_2\delta_2 +\eta_3\delta_3].
\end{equation}

The corresponding posterior density will be of the same form, with adjusted hyperparameters, i.e.,

\begin{equation}\label{post-dens}
f(\underline{\delta}|\underline{t}) \propto [\widetilde{K}(\delta_1,\delta_2,\delta_3)]^{\eta_0+n} \exp[(\eta_1+t_1) \delta_1 +(\eta_2+t_2)\delta_2 +(\eta_3+t_3)\delta_3 ],
\end{equation}

\noindent where $t_1=\sum x_i, t_2=\sum y_i $ and $t_3=\sum x_i y_i.$

However, in the process of selecting the values of the hyperparameters, we note that the posterior density is proportional to the likelihood of a sample of size $n_P=\eta_0+n)$ with sufficient statistics $t_{i,P}=\eta_i+t_i$, $i=1,2,3.$

Now, in order to make a sensible choice for the four hyperparameters, we will rely on the fact that our informed expert has had past experience with data very similar  to the current data set.
We ask him for a typical value for observed $X$'s which will be denoted by $v_1$, a typical value for the $Y$'s  to be denoted by $v_2$ and a typical value for the $XY$'s to be denoted by $v_3$.
Then we ask for a number or index to indicate how confident he is about the three typical values. Denote this by $n^{*}$. This can alternatively be viewed as being a consequence of having observed an ``imaginary'' sample of size $n^{*}$  with corresponding sufficient statistics
$$\sum_{i=1}^{n^{*}}x_i=n^{*}v_1,  \ \sum_{i=1}^{n^{*}}y_i=n^{*}v2, \ \sum_{i=1}^{n^{*}}x_iy_i=n^{*}v_3.$$

\bigskip

Based on this information we choose as our four hyperparameters $\eta_0=n^{*}$, $\eta_i=n^{*}v_i, \ \ i=1,2,3.$

\bigskip

We can rewrite the posterior density as a function  of original $\lambda_i$'s, if we wish. If we do so, it will be the same as the log-likelihood given in Eq. (5.4) with suitably revised values for $n,t_1,t_2$ and $t_3$. So, if we decide to use the posterior mode to estimate the $\lambda_i$'s, we can apply our iterative scheme to find the location of the mode.

\section{ A simulation study}
Let us assume that the confidence index provided by our informed expert is $n^{*}=12$, a small value, indicating that the expert is not at all sure about the values  $v_1=5,$ $v_2=4,$ $v_3=6$, that are provided. Then, as per the suggestion made earlier, we have the following suggested values for the hyperparameters $\eta_0=5$  $\eta_1=60,$ $\eta_2=48,$  $\eta_2=72.$   Next, the marginal posteriors can be obtained as (proportional to)

\begin{itemize}
\item $$\Pi_{1}\left(\delta_{1}| \vec{t^{*}}\right)\propto \exp[(\eta_1+t_1) \delta_1]\int_{-\infty}^{\infty} \int_{-\infty}^{0} [ \widetilde{K}(\delta_1,\delta_2,\delta_3)]^{\eta_0+n} \exp\left[(\eta_2+t_2)\delta_2 +(\eta_3+t_3)\delta_3\right] 
d\delta_{2} d\delta_{3}.$$

\item $$\Pi_{2}\left(\delta_{2}| \vec{t^{*}}\right)\propto \exp[(\eta_2+t_2) \delta_2]\int_{-\infty}^{\infty} \int_{-\infty}^{0} [ \widetilde{K}(\delta_1,\delta_2,\delta_3)]^{\eta_0+n} \exp\left[(\eta_1+t_1)\delta_1 +(\eta_3+t_3)\delta_3\right]d\delta_{1} d\delta_{3}.$$

\item $$\Pi_{3}\left(\delta_{3}| \vec{t^{*}}\right)\propto \exp[(\eta_3+t_3) \delta_3]\int_{-\infty}^{\infty}\int_{-\infty}^{\infty} [ \widetilde{K}(\delta_1,\delta_2,\delta_3)]^{\eta_0+n} \exp\left[(\eta_1+t_1)\delta_1 +(\eta_2+t_2)\delta_2\right]d\delta_{1} d\delta_{2}.$$

\end{itemize}

\bigskip

If instead we wish to use the posterior expectations of the $\lambda_i$'s as our estimates we will need to use numerical integration as follows.

\begin{itemize}
\item For $\lambda_1$,

$$E\left(\lambda_1|\underline{t}\right)=\frac{\int_{-\infty}^{\infty}\int_{-\infty}^{\infty}\int_{-\infty}^{0}e^{\delta_1} [\widetilde{K}(\delta_1,\delta_2,\delta_3)]^{\eta_0+n} \exp[(\eta_1+t_1) \delta_1 +(\eta_2+t_2)\delta_2 +(\eta_3+t_3)\delta_3 ] \ d\delta_1 \ d\delta_2 \ d\delta_3}{\int_{-\infty}^{\infty}\int_{-\infty}^{\infty}\int_{-\infty}^{0} [\widetilde{K}(\delta_1,\delta_2,\delta_3)]^{\eta_0+n} \exp[(\eta_1+t_1) \delta_1 +(\eta_2+t_2)\delta_2 +(\eta_3+t_3)\delta_3 ] \ d\delta_1 \ d\delta_2 \ d\delta_3}.$$

\item For $\lambda_2$,

$$E\left(\lambda_2|\underline{t}\right)=\frac{\int_{-\infty}^{\infty}\int_{-\infty}^{\infty}\int_{-\infty}^{0}e^{\delta_2} [\widetilde{K}(\delta_1,\delta_2,\delta_3)]^{\eta_0+n} \exp[(\eta_1+t_1) \delta_1 +(\eta_2+t_2)\delta_2 +(\eta_3+t_3)\delta_3 ] \ d\delta_1 \ d\delta_2 \ d\delta_3}{\int_{-\infty}^{\infty}\int_{-\infty}^{\infty}\int_{-\infty}^{0} [\widetilde{K}(\delta_1,\delta_2,\delta_3)]^{\eta_0+n} \exp[(\eta_1+t_1) \delta_1 +(\eta_2+t_2)\delta_2 +(\eta_3+t_3)\delta_3 ] \ d\delta_1 \ d\delta_2 \ d\delta_3}.$$

\item For $\lambda_3$,

$$E\left(\lambda_3|\underline{t}\right)=\frac{\int_{-\infty}^{0}\int_{-\infty}^{\infty}\int_{-\infty}^{\infty}e^{\delta_3} [\widetilde{K}(\delta_1,\delta_2,\delta_3)]^{\eta_0+n} \exp[(\eta_1+t_1) \delta_1 +(\eta_2+t_2)\delta_2 +(\eta_3+t_3)\delta_3 ] \ d\delta_1 \ d\delta_2 \ d\delta_3}{\int_{-\infty}^{\infty}\int_{-\infty}^{\infty}\int_{-\infty}^{0} [\widetilde{K}(\delta_1,\delta_2,\delta_3)]^{\eta_0+n} \exp[(\eta_1+t_1) \delta_1 +(\eta_2+t_2)\delta_2 +(\eta_3+t_3)\delta_3 ] \ d\delta_1 \ d\delta_2 \ d\delta_3}.$$

\end{itemize}

\noindent Note that higher order moments can also be obtained (via numerical methods, of course).  The choice of priors will have a significant impact on both bias and computational time.

\noindent We consider the posterior mean as the Bayes estimates for the parameters. We also provide 95\% credible intervals as a summary related to Bayesian estimation that are given in Table $6.1.$ We consider the following four different parameter settings:

\begin{itemize}
\item {\bf Choice 1:} $\delta_{1}=-2.5; \quad \delta_{2}=-1.3; \quad \delta_{3}=-0.25. $

\item {\bf Choice 2:} $\delta_{1}=-1.8; \quad \delta_{2}=-0.98; \quad \delta_{3}=-0.45. $

\item {\bf Choice 3:} $\delta_{1}=1.58; \quad \delta_{2}=1.87; \quad \delta_{3}=-0.55. $

\item {\bf Choice 4:} $\delta_{1}=0.46; \quad \delta_{2}=0.92; \quad \delta_{3}=-0.65. $
\end{itemize}

\begin{sidewaystable}
\centering
\small
\caption{Posterior summary for the BPC model under the conjugate prior assumption}
\begin{tabular}{|l|l|l|l|l|l|l|l|l}
\hline
\multirow{4}{*}{Parameter choices}&
      \multicolumn{2}{c}{$\widehat{\lambda_{1}}$} &
      \multicolumn{2}{c}{$\widehat{\lambda_{2}}$} &
     \multicolumn{2}{c}{$\widehat{\lambda_{3}}$}\\
& Posterior mean &95\% HPD &Posterior mean &95\% HPD&Posterior mean &95\% HPD\\ 
\hline
Choice 1& 0.0751 &(0.0386, 2.9921) & 0.2611& (0.0767, 1.8154)& 0.7718 &( 0.5018, 0.8541)\\
\hline
 Choice 2& 0.1725& (0.1277, 1.0568) & 0.3598& (0.1429, 1.3422)& 0.6389& (0.3479,0.6817)\\
\hline
 Choice 3& 4.8695 &(1.076, 6.3756) & 1.932 &(1.1921, 3.8764)& 0.5521&( 0.5040, 0.9483)\\
\hline
 Choice 4&1.5688 &(1.389, 5.0218) & 2.487 &(1.597, 3.4856)& 0.5127 &(0.4082, 0.7527)\\ 
\hline
\end{tabular}
\end{sidewaystable}

\subsection{ Bayesian analysis with locally uniform priors}
In this case, we consider the following locally uniform priors for the three parameters which are as follows:

\begin{itemize}
\item $\Pi(\delta_1)\propto 1, \quad \text{for} \quad -\infty<\delta_1<\infty.$

\item $\Pi(\delta_2)\propto 1, \quad \text{for} \quad -\infty<\delta_2<\infty,$ $\quad$  and 

\item $\Pi(\delta_3)\propto 1, -\infty<\delta_3<0.$

\end{itemize}

\noindent If we, before observing the imaginary sample, assume  that the parameters had a flat joint prior  (that are given above), then the posterior, after observing the imaginary sample would be just the likelihood of the imaginary sample, i.e.,

\begin{equation}
L^{*}(\underline{\delta})\propto [ \widetilde{K}(\delta_1,\delta_2,\delta_3)]^{n{*}} \exp\left[\delta_1 n^{*}v_1+\delta_2 n^{*}v_2+ \delta_3n^{*}v_3\right].
\end{equation}

It is this posterior that we will use for a prior for the real data set. Therefore, the resulting posterior combining the data likelihood given in Eq. (5.2)  with the prior given in Eq. (6.2) will be 

\begin{equation}
\Pi\left(\underline{\delta}|\vec{v}, \vec{t}\right)\propto [ \widetilde{K}(\delta_1,\delta_2,\delta_3)]^{n{*}+n} 
\exp\left[\delta_1 \left(n^{*}v_1+t_1\right)+\delta_2 \left(n^{*}v_2+t_2\right) \delta_3\left(n^{*}v_3+t_3\right)\right].
\end{equation}

\noindent Subsequently, the posterior means for the parameters are obtained which  will be 

\begin{itemize}
 \item 
 \small
 \begin{align*}
&E\left[\lambda_{1}|\vec{v}, \vec{t}\right]\notag\\
&=\frac{\int_{-\infty}^{\infty}\int_{-\infty}^{\infty} \int_{-\infty}^{0} [ \widetilde{K}(\delta_1,\delta_2,\delta_3)]^{n^{*}+n}\bigg(
\exp\left[\delta_1 \left(n^{*}v_1+t_1+1\right)\delta_2 \left(n^{*}v_2+t_2\right)+ \delta_3\left(n^{*}v_3+t_3\right)\right]\bigg)d\delta_{1}d\delta_{2}d\delta_{3}}
 {\bigg(\int_{-\infty}^{\infty}\int_{-\infty}^{\infty}\int_{-\infty}^{0} [ \widetilde{K}(\delta_1,\delta_2,\delta_3)]^{n{*}+n} 
\exp\left[\delta_1 \left(n^{*}v_1+t_1\right)+\delta_2 \left(n^{*}v_2+t_2\right) \delta_3\left(n^{*}v_3+t_3\right)\right] d\delta_{1}d\delta_{2}d\delta_{3}\bigg)}
\end{align*}

\item 
 \small
 \begin{align*}
&E\left[\lambda_{2}|\vec{v}, \vec{t}\right]\notag\\
&=\frac{\int_{-\infty}^{\infty}\int_{-\infty}^{\infty} \int_{-\infty}^{0} [ \widetilde{K}(\delta_1,\delta_2,\delta_3)]^{n^{*}+n}\bigg(
\exp\left[\delta_1 \left(n^{*}v_1+t_1\right)\delta_2 \left(n^{*}v_2+t_2+1\right)+ \delta_3\left(n^{*}v_3+t_3\right)\right]\bigg)d\delta_{1}d\delta_{2}d\delta_{3}} {\int_{-\infty}^{\infty}\int_{-\infty}^{\infty}\int_{-\infty}^{0} [ \widetilde{K}(\delta_1,\delta_2,\delta_3)]^{n{*}+n} 
\exp\left[\delta_1 \left(n^{*}v_1+t_1\right)+\delta_2 \left(n^{*}v_2+t_2\right) \delta_3\left(n^{*}v_3+t_3\right)\right] d\delta_{1}d\delta_{2}d\delta_{3}}.
\end{align*}

\item  
 \small
 \begin{align*}
&E\left[\lambda_{3}|\vec{v}, \vec{t}\right]\notag\\
&=\frac{\int_{-\infty}^{\infty}\int_{-\infty}^{\infty} \int_{-\infty}^{0} [ \widetilde{K}(\delta_1,\delta_2,\delta_3)]^{n^{*}+n}\bigg(
\exp\left[\delta_1 \left(n^{*}v_1+t_1\right)\delta_2 \left(n^{*}v_2+t_2\right)+ \delta_3\left(n^{*}v_3+t_3+1\right)\right]\bigg)d\delta_{1}d\delta_{2}d\delta_{3}}
{\int_{-\infty}^{\infty}\int_{-\infty}^{\infty}\int_{-\infty}^{0} [ \widetilde{K}(\delta_1,\delta_2,\delta_3)]^{n{*}+n} 
\exp\left[\delta_1 \left(n^{*}v_1+t_1\right)+\delta_2 \left(n^{*}v_2+t_2\right) \delta_3\left(n^{*}v_3+t_3\right)\right] d\delta_{1}d\delta_{2}d\delta_{3}}.
\end{align*}

\end{itemize}

\smallskip

\noindent In this case, we use the same set of four different choices of the model parameters listed earlier for the simulation study.

\begin{sidewaystable}
\centering
\small
\caption{Posterior summary for the BPC model under the non-informative prior assumption}
\begin{tabular}{|l|l|l|l|l|l|l|l|l}
\hline
\multirow{4}{*}{Parameter choices}&
      \multicolumn{2}{c}{$\widehat{\lambda_{1}}$} &
      \multicolumn{2}{c}{$\widehat{\lambda_{2}}$} &
     \multicolumn{2}{c}{$\widehat{\lambda_{3}}$}\\
& Posterior mean &95\% HPD &Posterior mean &95\% HPD&Posterior mean &95\% HPD\\ 
\hline
Choice 1& 0.0725 &(0.03148, 2.8066) & 0.2432& (0.0831, 7.7160)& 0.7795 &( 0.5204, 0.8503)\\
\hline
 Choice 2& 0.1676& (0.1125, 3.2384) & 0.3113& (0.1316, 2.3809)& 0.6492& (0.2841, 0.6756)\\
\hline
 Choice 3& 4.9258 &(2.7871, 11.2054) & 6.6925 &(2.8171, 9.4934)& 0.5559 &( 0.4904, 0.9525)\\
\hline
 Choice 4& 3.2546 &(1.295, 5.2314) & 3.485 &(1.4782, 5.4218)& 0.5071 &(0.3929, 0.7459)\\ 
\hline
\end{tabular}
\end{sidewaystable}

\section{Bayesian inference using posterior mode}

If we re-write the joint posterior in Eq. (2.4) in terms of the original $\lambda_{i}'$ s, the expression will be [under the conjugate prior set-up]

\begin{equation}
 \label{post-dens1}
f(\underline{\lambda}|\underline{t}) \propto [K(\lambda_1,\lambda_2,\lambda_3)]^{\eta_0+n} \lambda^{\eta_1+t_1}_{1}\lambda^{\eta_2+t_2}_{2}\lambda^{\eta_3+t_3}_{3}.
\end{equation}

%\begin{equation}
% \label{post-dens2}
%\Pi(\lambda_{1}|\underline{t})
% \propto\lambda^{\eta_1+t_1}_{1}\int_{0}^{\infty}\int_{0}^{1} [K(\lambda_1,\lambda_2,\lambda_3)]^{\eta_0+n} \lambda^{\eta_2+t_2}_{2}\lambda^{\eta_3+t_3}_{3}d\lambda_{2}d\lambda_{3}
%\end{equation}

%Similarly, one can find the expression for the other two marginal posteriors. 
 Next, it is straightforward to to find the posterior mode of $\left(\lambda_1, \lambda_2,\lambda_3 \right)$ using Newton-Raphson and to obtain approximate posterior standard deviations of $\left(\lambda_1, \lambda_2,\lambda_3 \right)$ using the second derivative matrix of the log posterior evaluated at the mode. However, we only report the posterior mode values.

\smallskip

For the simulation study, we select the same set of parameter choices as in the case of MLE. A random sample of size $n=100$ is drawn from the joint distribution. 
\begin{itemize}
\item[(a)] Choice 1: $\lambda_{1}= 2$, $\lambda_{2}=2.5$ and  $\lambda_{3}=0.35.$
\item[(b)] Choice 2: $\lambda_{1}= 1.75$, $\lambda_{2}=3.25$ and  $\lambda_{3}=0.45.$
\item[(c)] Choice 3: $\lambda_{1}= 2.5$, $\lambda_{2}=1.5$ and  $\lambda_{3}=0.55.$
\item[(d)] Choice 4: $\lambda_{1}= 3.5$, $\lambda_{2}=4$ and  $\lambda_{3}=0.75.$
\end{itemize}

For the conjugate prior set-up, we consider the following {\color{blue} values} of the hyperparameters: $\eta_{0}=1.23, \quad \eta_1=2.325, \quad \eta_2=3.25, \quad \eta_{3}=2.528.$
We report the location of the posterior modes  of the posterior  as a summary related to Bayesian estimation that are given in Table $7.1.$

 \begin{table}
\centering
\small
\caption{Posterior modes for the BPC model under the conjugate prior assumption}
\begin{tabular}{|l|l|l|l|l|l|l|l|l}
\hline
\multirow{4}{*}{Parameter choices}&
      \multicolumn{1}{c}{$\widehat{\lambda_{1}}$} &
      \multicolumn{1}{c}{$\widehat{\lambda_{2}}$} &
     \multicolumn{1}{c}{$\widehat{\lambda_{3}}$}\\
& Posterior mode &Posterior mode& Posterior mode\\ 
\hline
Choice 1& 2.131  & 2.522& 0.315\\
\hline
 Choice 2& 1.784  & 1.5783 & 1.467 \\
\hline
 Choice 3& 2.539  & 1.487  & 0.583 \\
\hline
 Choice 4& 3.601 & 3.926  & 0.743\\ 
\hline
\end{tabular}
\end{table}

\section{Real-data application}
To illustrate the feasibility of the proposed two Bayesian approaches in the preceding section, we consider the data which is originally due to Aitchison and Ho (1989). This data set has also been studied independently by Lee et al. (2017) and Ghosh et al. (2021).  For pertinent details on this particular data set and the applicability of the bivariate Poisson conditionals distribution as a reasonable fit for this data set, see Ghosh et al. (2021). In this subsection, we  re-analyze this dataset under the Bayesian paradigm.  \\
 
 \noindent Next, for the Bayesian analysis, we make a note of the following:
 
 \begin{itemize}
 \item For the conjugate prior set-up, we consider the following  values of the hyperparameters: $\eta_{0}=1.41, \quad \eta_1=2.325, \quad \eta_2=3.25, \quad \eta_{3}=2.528.$
 \item For the locally uniform prior set-up, we consider the joint prior as given in Eq. (6.1).
 \end{itemize}
 
\noindent  The parameter estimates (posterior mean, highest  posterior density interval) under both the conjugate prior and the locally uniform priors are provided in Table $8.1.$
 
\begin{sidewaystable}
\centering
\small
\caption{Goodness of fit summary for the Lens data under the BPC model )}
\begin{tabular}{|l|l|l|l|l|l|l|l|l}
\hline
\multirow{2}{*}{Parameter choices}&
      \multicolumn{2}{c}{$\widehat{\lambda_{1}}$} &
      \multicolumn{2}{c}{$\widehat{\lambda_{2}}$} &
     \multicolumn{2}{c}{$\widehat{\lambda_{3}}$}\\
& Posterior mean &95\% HPD &Posterior mean &95\% HPD&Posterior mean &95\% HPD\\ 
\hline
Conjugate prior set-up& 1.8500 &(1.3832, 3.6052) & 2.1699& (1.7633, 6.400)& 0.9600 &( 0.5574, 0.9832)\\
\hline
 Locally uniform prior set-up& 1.8650& (1.2926, 4.1516) & 2.1878& (1.6507,6.8579)& 0.9574& (0.3378, 1.0211)\\
\hline
\end{tabular}
\end{sidewaystable}

\smallskip

\noindent From the Table $6.1,$  it appears that the parameter estimates obtained with the conjugate prior choice closely matches  the MLE estimates obtained using  the copula as discussed in Ghosh et al. (2021). Under the flat prior set-up, the length of 95\% HPD intervals are slightly wider  as can be observed from Table $8.1,$ second row---third, fifth and the seventh column values.

 \section{Conclusion}
Modeling  of bivariate paired count data is an open problem because of the inadequate class of bivariate discrete distributions, which if available, might explain the true dependence structure effectively. In this  paper,  we focus on the  classical (using  an iterative approach) and Bayesian inference for a bivariate discrete probability distribution for which both the conditionals belong to an univariate Poisson distribution with appropriate parameters, and the distribution is described by Arnold et al. (1999), which will always have negative correlation, except in the independent case.  In this paper, we have discussed an alternative iterative algorithm for the maximum likelihood method under the frequentist set-up which has a striking advantage that we don't need any maximizing/optimizing root finding subroutines and which can be  implemented in any programming environment via some user defined package(s). 
On the Bayesian inferential aspect, both the conjugate and the locally uniform prior set-up have been assumed. While a conjugate prior set-up is quite natural for the joint p.m.f. of the form as given in (5.1),  it is challenging to find a conjugate prior in such a scenario from a real-world perspective.
  A full scale study under both the classical and Bayesian paradigm for a multivariate Poisson conditional distribution can be considered from a real-life perspective where such a model will be useful. We did not pursue this problem here as it is beyond the scope of this paper.

  \section*{Disclosure Statement}
  The authors do not have a competing interest.

\end{document}